\documentclass[prx, twocolumn,preprintnumbers]{revtex4-1}

\usepackage{amsmath,amssymb,mathrsfs}
\usepackage{subfigure}
\usepackage[dvips]{graphicx}
\usepackage{color}
\usepackage{bbold}

\def\be{\begin{equation}}
\def\ee{\end{equation}}
\def\bea{\begin{eqnarray}}
\def\eea{\end{eqnarray}}
\def\nn{\nonumber}

\usepackage[utf8x]{inputenc}

\begin{document}

\title{Cyclotron dynamics of interacting bosons in artificial magnetic fields }
\author{Xiaopeng Li}
\author{S. Das Sarma}
\affiliation{Condensed Matter Theory Center and Joint Quantum Institute, Department of Physics, 
University of Maryland, College Park, MD 20742-4111, USA}

\begin{abstract}
 We study {theoretically} quantum dynamics of interacting bosons in artificial magnetic fields as engineered in recent 
ultracold atomic experiments, where quantum cyclotron orbital motion has been observed. With exact numerical simulations and 
perturbative analyses, we find that interactions induce damping in the cyclotron motion. 
The damping time is found to be dependent on {interaction and tunneling strengths} monotonically, while its dependence 
on magnetic flux is non-monotonic. 
Sufficiently strong interactions would render bosons dynamically localized inhibiting the cyclotron motion. 
 {The damping predicted by us can be construed as an interaction-induced quantum decoherence of the cyclotron motion.}
\end{abstract}

\date{\today} 
\maketitle 

\section{Introduction} 
Cyclotron orbits and Landau levels formed by electrons moving in magnetic fields play an 
essential role in the emergence of several novel phenomena 
in solid state {systems}. Semiclassical cyclotron orbital motion 
in two dimensional electron gas gives rise to the Hall conductance 
{and then eventually to quantized Hall conductance in high enough magnetic fields. 
In quantum Hall insulators, chiral edge states mediating 
dissipationless edge current~\cite{1981_Langhlin_PRB,1982_TKNN} can be understood as quantum cyclotron orbits 
bounded by the edges in a  {strip-like geometry.} }
{In the current work, we study the cyclotron motion of {\it interacting} 
bosonic neutral atoms in an optical lattice subjected to artificial  gauge fields 
which act like effective external magnetic fields in the lattice leading to novel physics~\cite{2013_Galitski_NatReview}. }

{Ultracold atomic gases confined in optical lattices, because of their unprecedented  controllability,  
allow for quantum simulations of various lattice Hamiltonians,} 
e.g., Bose-Hubbard models~\cite{1989_Fisher_PRB,1998_Zoller_PRL}, 
where both equilibrium many-body physics~\cite{2002_Greiner_Nature,2005_Bloch_Interferometry} and non-equilibrium 
dynamics have been {extensively studied theoretically~\cite{2011_Anatoli_NonEq_RMP} 
and experimentally}~\cite{2002_Greiner_Revival,2012_Bloch_lightcone}. 
Recent experiments {created}
a two dimensional square lattice {pierced by} magnetic flux {(``artificial gauge fields'')} by engineering laser assisted 
tunneling~\cite{2011_Bloch_PRL,2013_Bloch_Harper,2013_Ketterle_Harper}. 
{Due to non-trivial Berry curvatures {in such a system,} 
charge neutral atoms, e.g. $^{87}$Rb, loaded into this flux lattice 
behave like ``charged'' bosons  experiencing strong magnetic fields, and the consequent effective Lorentz force 
results in atomic cyclotron motion,  which has been {experimentally} observed~\cite{2013_Bloch_Harper}. 
While these {interesting} experimental developments 
are largely motivated {by considerations of observing fascinating new} 
equilibrium many-body phases such as atomic quantum spin Hall
insulators~\cite{2009_Galitski_TopoOL_PRA,2010_DasSarma_TopoOL_PRA,2010_Goldman_TIOL_PRL,2011_Cooper_TI_PRL,2013_Ketterle_QSH,2013_Liu_TopOL_PRL,2013_Zhao_TI}, 
 {the observed non-equilibrium cyclotron dynamics of bosons itself~\cite{2013_Bloch_Harper} 
is extremely interesting and requires theoretical understanding. 
In particular, the interaction effect on the dynamical cyclotron motion is obviously of great interest, and is the main topic 
of study in the current work.}

In this article, we study cyclotron dynamics of interacting bosons with 
both exact numerical simulations and perturbative analyses.  
Weak interactions are found to induce damping effects (i.e., quantum decoherence) in the dynamics.
We find that while the damping time (or the decoherence time) monotonically decreases 
with increasing tunneling and interaction,  it has non-monotonic behavior with varying magnetic 
flux. 
{With the perturbative analyses, the damping effect 
is attributed to specific scattering processes, and such physics is established to be generic 
for  interacting bosons in artificial gauge fields, i.e., not relying on the model Hamiltonian used 
in our numerical simulations.} 
With sufficiently strong interactions, cyclotron dynamics is completely 
suppressed and  the bosons form a dynamically localized state, analogous to self-trapping effects observed 
in Bose-Einstein condensates in double-well potentials~\cite{1997_Milburn_SelfTrapping,1997_Smerzi_SelfTrapping,
2005_Oberthaler_SelfTrapping,2006_WuBiao_SelfTrapp_PRA}. 
Our finding of the cyclotron damping effect suggests importance of interactions and many-body physics 
in quantum transport of bosons in artificial gauge fields, 
which is of great interest in recent atomic gases~\cite{2009_Lin_Nat,2011_Lin_Spielman_Nature,2013_Ketterle_QSH}. 
Quantum simulations of this damping effect in such controllable systems would help understand 
relaxations in Hall transport experiments in complex electronic materials where the decoherence could be attributed 
to various origins.

\begin{figure}[htp]
\begin{center}
\includegraphics[angle=0, width=\linewidth]{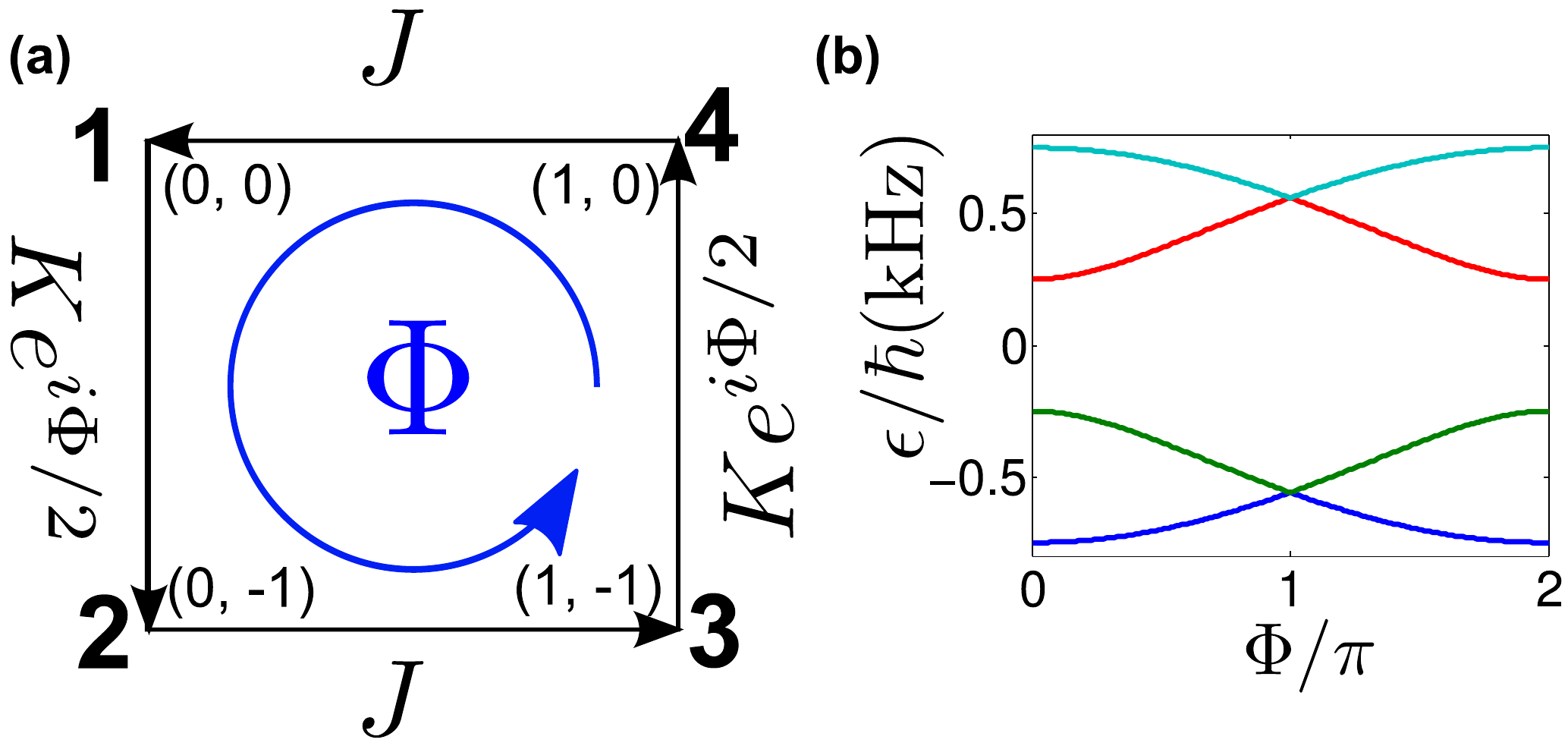}
\end{center}
\caption{  An optical lattice plaquette with magnetic flux $\Phi$ and its spectra. (a) The plaquette system 
where the sites $1$, $2$, $3$ and $4$ are located at 
$(0,0)$, $(0, -1)$, $(1, -1)$, and $(1,0)$ in our coordinate choice. (b) The single-particle 
energy spectra with varying $\Phi$. 
} 
\label{fig:system}	
\end{figure}

\section{ System and model Hamiltonian} 
In the  experimental setup to observe the cyclotron motion~\cite{2013_Bloch_Harper}, 
a plaquette of four lattice sites is isolated in a two dimensional 
optical lattice by suppressing inter-plaquette tunnelings with superlattice techniques. 
To study the cyclotron motion, we look at one 
isolated plaquette threaded by magnetic flux as illustrated in Fig.~\ref{fig:system}. 
The model Hamiltonian describing bosons loaded into this plaquette is 
{($\hbar=1$ throughout)}
\bea
H &=& H_0 + V \nn \\ 
H_0 &=&  -K \left[ e^{i\Phi/2} ( b_2 ^\dag b_1 + b_4 ^\dag b_3 ) + h.c.\right] \nn \\
&&-J \left[ b_3 ^\dag b_2 + b_4 ^\dag b_3 + h.c. \right]  \nn \\
V &=& \frac{U}{2} \sum_j b_j ^\dag b_j ^\dag b_j b_j, 
\label{eq:Ham}
\eea 
where $b_{j}$ is a bosonic annihilation operator for the $j$th site. 
This engineered Hamiltonian connects to charged bosons in magnetic fields 
through Peierls substitution~\cite{1976_Hofstadter_PRB,2012_Spielman_PRL}.
The tunneling 
strength $J$ is  fixed in our calculation to be $0.5\times 2\pi$kHz 
following the experimental situation. 
The free part of {the } Hamiltonian can be written as 
$H_0 = \sum_{j j'} {\cal H} ^{(0)} _{j j'} b_j ^\dag b_{j'}$, 
with ${\cal H} ^{(0)}$ the single-particle Hamiltonian matrix. 
We study the quantum dynamics assuming an initial state
\bea 
|\Psi\rangle =\frac{1}{\sqrt{N!}} \left[\psi ^\dag \right] ^ N |0\rangle, 
\label{eq:initstate}
\eea 
with $\psi^\dag = \frac{1}{\sqrt{2}} \left( b_3 ^\dag + b_4 ^\dag \right)$, 
which describes $N$ bosons prepared in a superposed state of sites $3$ and $4$. 
The physics described here is otherwise robust against  the choice of $\psi$ as long as it is not 
fine-tuned.

To characterize the cyclotron motion, the time dependent occupation numbers $n_j$ and an average 
position vector $\overline{\vec{X}} (t) = (x(t), y(t))$ are defined as 
\bea 
n_j (t) &=& \frac{1}{N} \langle \Psi(t) | b_j ^\dag b_j |\Psi(t)\rangle, \\
\overline{\vec{X}} (t) &=& \sum_j \vec{R}_j n_j (t) , 
\eea 
where $|\Psi(t)\rangle$ is the time evolved many-body state, and $\vec{R}_j$ is the position of the $j$-th 
site (see Fig.~\ref{fig:system}(a)). The initial state  is not an eigenstate of 
the Hamiltonian and is thus not {stationary}. 
For non-interacting bosons, we have 
$
|\Psi (t)  \rangle = \frac{1}{ \sqrt{N!}} \left [ \psi^\dag (t) \right] ^N |0\rangle, 
$
with 
$\psi ^\dag (t) = e^{-iH_0 t} \psi ^\dag e^{iH_0 t} =  \sum_j \psi_j (t) b_j ^\dag, $
where the coefficients $\psi_j (t)$ 
are determined by the single-particle Schr\"odinger equation 
$i\partial_t \psi_j (t)  = \sum_{j'} {\cal H}^{(0)} _{j j'} \psi_{j'} (t)$. 
In this state, bosons actually rotate in the plaquette when the engineered magnetic flux is non-zero 
(see Fig.~\ref{fig:cyclotron_Flux}), which is a 
quantum analogue of  classical charged particles 
moving in a magnetic field, and this quantum cyclotron motion is undamped. 
The density inhomogeneity among the four sites 
{ oscillates without any relaxation.}  
One useful quantity in this dynamical process is {the} occupation fraction 
$P_\psi (t) = N_\psi (t) / N$  with 
 $$
 N_\psi (t) = \langle \Psi(t) | \psi^\dag (t) \psi (t)|\Psi(t)\rangle, 
 $$
 where 
$N_\psi(t)$ can be thought as the occupation number of the initially occupied single-particle {mode}  
$\psi (t)$. 
Although 
the quantum state  is fully dynamical and involves fast oscillations {on} the tunneling time scale, $J^{-1}$,  
(around one millisecond),  non-interacting bosons {remain} in the single-particle state 
$\psi(t)$, and  the occupation fraction $P_\psi(t)$ remains {unity}, 
{indicating a perfectly coherent bosonic cyclotron motion in the non-interacting optical lattice.} 

\begin{figure}[htp]
\begin{center}
\includegraphics[angle=0, width=.8\linewidth]{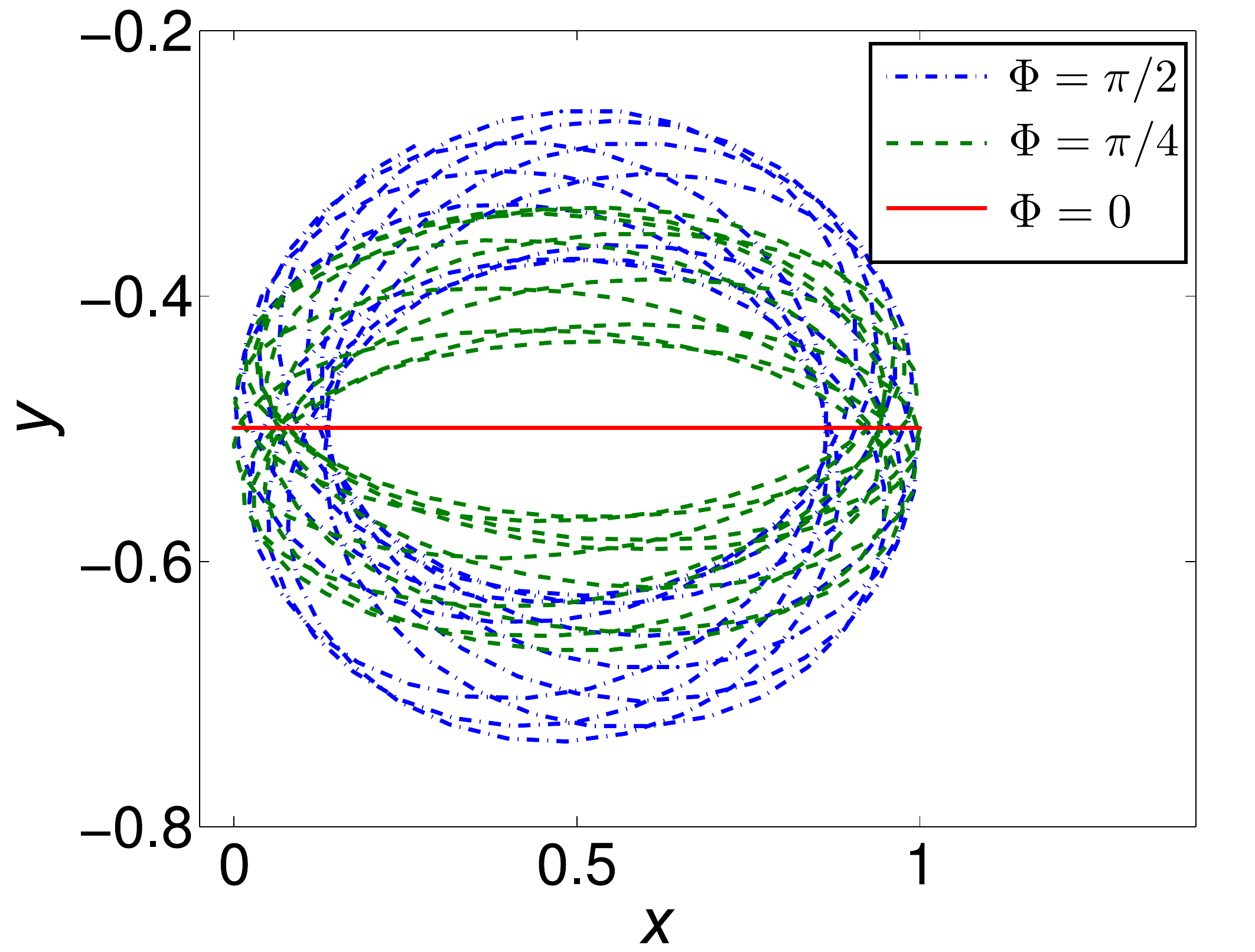}
\end{center}
\caption{Cyclotron motion of non-interacting bosons with various magnetic flux. 
Bosons are circulating in the plaquette with finite magnetic flux ($\pi/2$ and $\pi/4$ in this plot), and the 
circular dynamics is a quantum analogue of cyclotron motion. 
For magnetic flux $\Phi =0$, the dynamics  cannot be identified as 
cyclotron motion.} 
\label{fig:cyclotron_Flux}	
\end{figure}

\begin{figure}[htp]
\begin{center}
\includegraphics[angle=0, width= \linewidth]{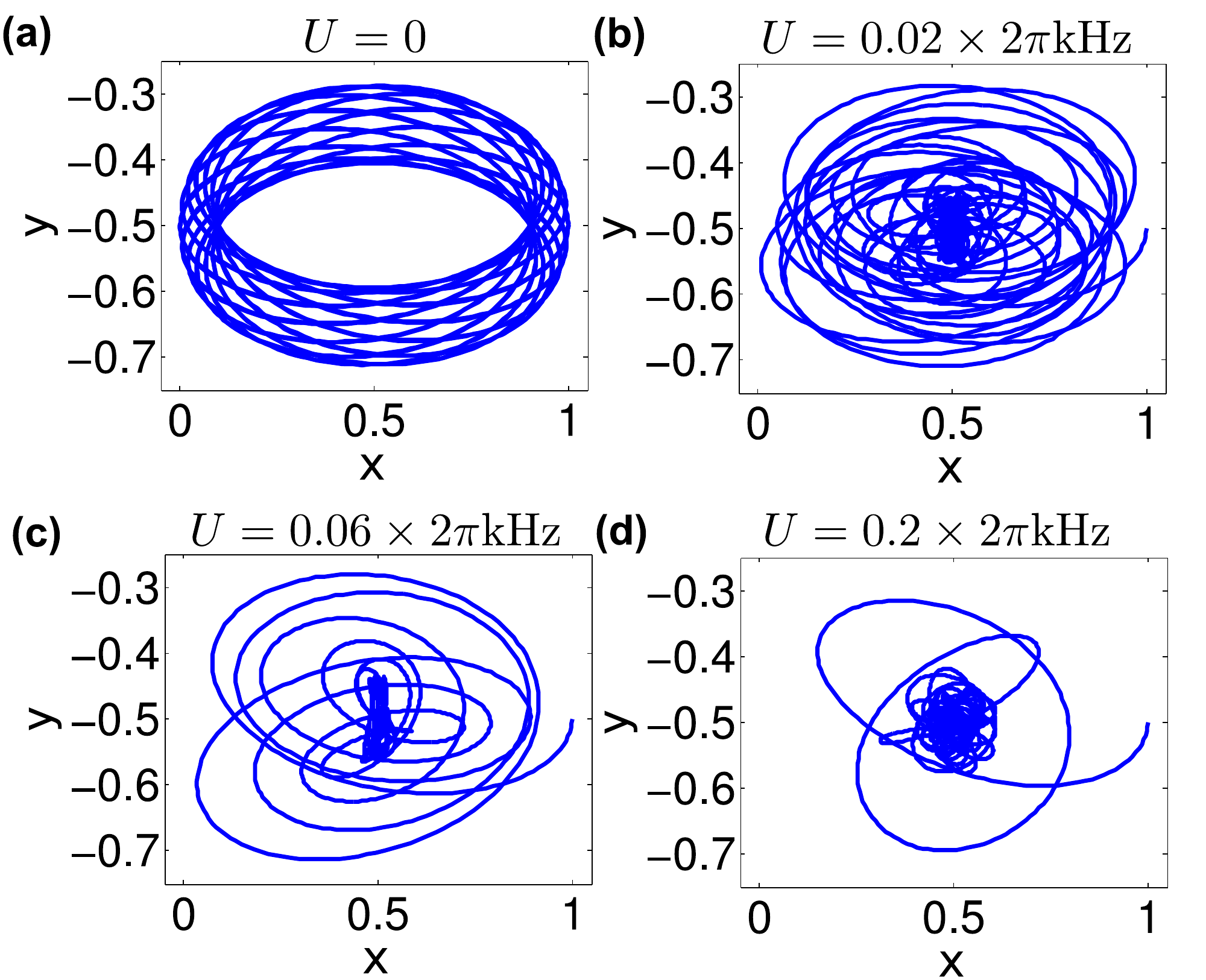}
\end{center}
\caption{ Cyclotron motion with various interaction strengths. 
The average position $(x(t), y(t))$ (see the main text) illustrates the rotation of bosons in the plaquette. 
(a) The cyclotron motion of non-interacting bosons, where damping does not occur. (b), (c) and (d) show 
the interacting case with varying interaction strength $U$. Interaction effects make $(x(t), y(t))$ collapse into the center of the plaquette after 
several periods of rotation. The periods it costs for the rotation to collapse decrease with increasing interaction 
strength. 
Here we use the parameters $K = 0.25 \times 2\pi$kHz, and $\Phi = 0.735 \times \pi/2$ 
as realized in the experiment~\cite{2013_Bloch_Harper}.  } 
\label{fig:cyclotronmotion}	
\end{figure}

\section{ Weakly interacting bosons} 
\subsection{Numerical simulations}
We first simulate the dynamics (Figs.~\ref{fig:cyclotronmotion},\ref{fig:depletionvsdamping})  
with an  exact treatment of  the many-body Schr\"odinger equation 
$i\partial_t |\Psi (t) \rangle = H |\Psi(t) \rangle$, 
 where the Hamiltonian $H$ and the time evolved state $|\Psi (t) \rangle$ are represented in a complete basis 
$$
|M_1 M_2 M_3 M_4 \rangle = \prod_j \frac{1}{\sqrt{ M_j !}} \left( b_j ^{\dag}  \right) ^{M_j} |0\rangle.
$$    
In our numerical simulations, the total particle number is fixed to be $8$,  i.e., the mean filling is 
two particles per site. 
{In Fig.~\ref{fig:depletionvsdamping}, the cyclotron motion illustrated by 
oscillations in the {average} position $\overline{\vec{X} }(t) = (x(t), y(t))$ 
shows damping in the presence of repulsive interactions. 
After several \mbox{(quasi-)}periods of cyclotron motion, $ \overline{\vec{X} } (t)$ collapses to the regime around 
the center of the plaquette (Fig.~\ref{fig:cyclotronmotion}).  
In this case, the $\psi$-mode occupation fraction 
$P_\psi (t)$  no longer {remains unity}, nonetheless it still remains quasi-static, 
namely, does not exhibit fast oscillations.  
The damping of oscillation amplitudes in $\overline{\vec{X} } (t)$  is found to 
{coincide} with the decrease in $P_\psi (t) $ (Fig.~\ref{fig:depletionvsdamping}). }  
The damping of cyclotron motion is thus well captured by $P_\psi (t)$. 
Physically, the decrease in $P_\psi (t)$ is caused by interaction processes where bosons 
are scattered out of {their} originally occupied $\psi$ {mode}  
(this physical picture is borne out by {our perturbative analysis presented below}). 
Its coincidence with the cyclotron motion damping 
implies that the scattered bosons do not contribute to the cyclotron motion coherently,  {thus contributing to quantum decoherence. }

\begin{figure}[htp]
\begin{center}
\includegraphics[angle=0, width= .8\linewidth]{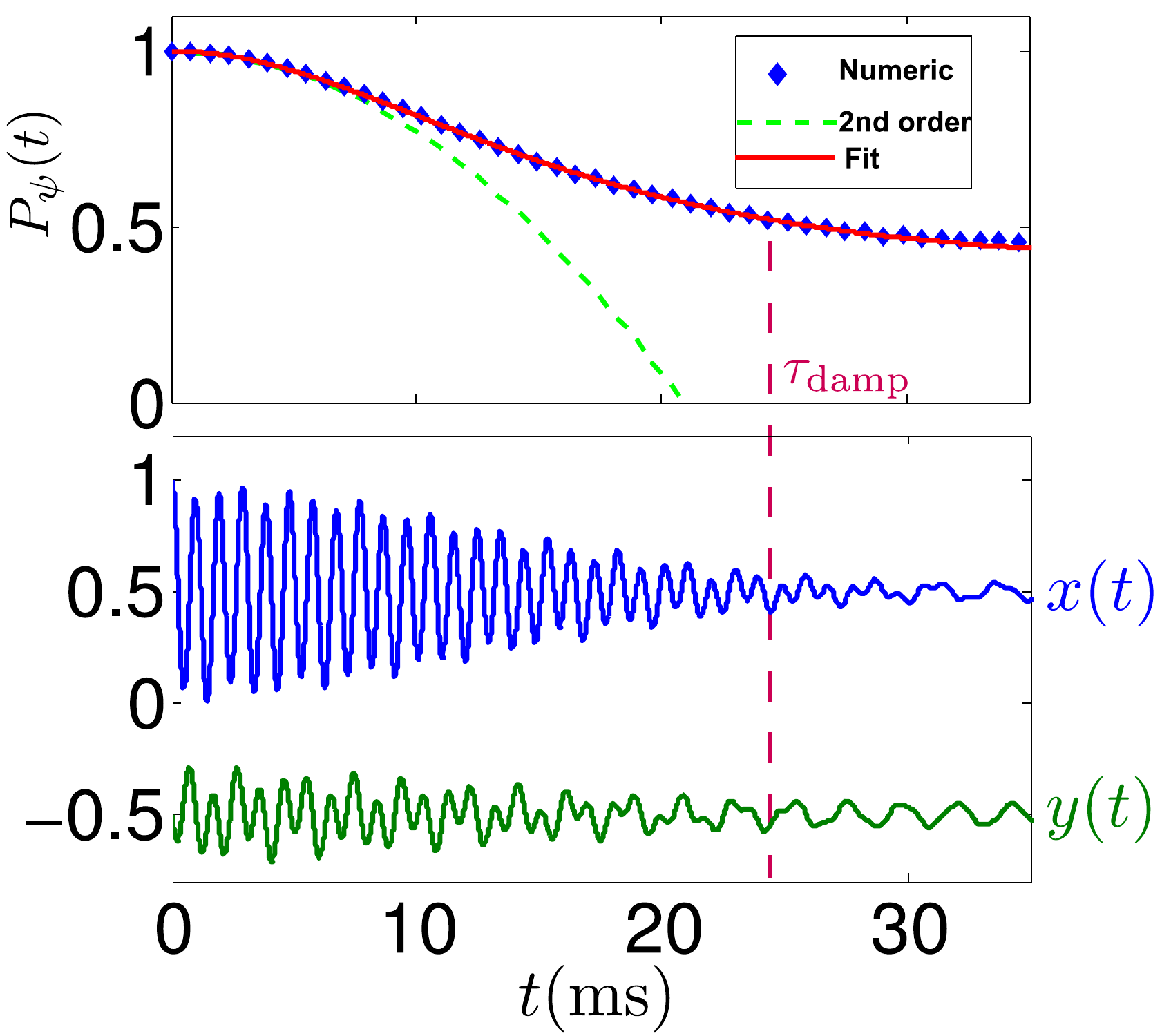}
\end{center}
\caption{{Damping of cyclotron motion and decay of the occupation fraction $P_\psi (t)$.} Top panel shows $P_\psi(t)$ obtained by $2$nd order 
perturbation theory and  by exact numerical simulations. The $2$nd order perturbation result agrees with numerics 
at short time as expected. In the intermediate regime where $t<\tau_{\rm damp}$,  $P_\psi(t)$ 
{is well described by  an empirical fit ${ P}_{\rm fit} (t)$ (Eq.~\eqref{eq:fitfunction}) } 
and the fitting error is negligible. Bottom panel shows the average position $(x(t), y(t))$.  Comparing two 
panels, the damping in $x(t)$ and $y(t)$ coincides with decrease of $P_\psi(t)$.  
{In this plot we use $U = 0.02 \times 2\pi {\rm kHz}$,  
$K = 0.25 \times 2\pi {\rm kHz} $ and $\Phi = 0.735 \times \pi/2$. } } 
\label{fig:depletionvsdamping}	
\end{figure}

The strength of damping can be quantified by a damping time (decoherence time) $\tau_{\rm damp}$ which 
we define to be the time it takes for half of the bosons in the $\psi$ mode to be scattered into other single-particle 
states, namely the time when $P_\psi (t=\tau_{\rm damp})$ reaches $1/2$. 
The damping time $\tau_{\rm damp}$ is found to be inversely proportional to the interaction strength 
when it is sufficiently weak. 
For the parameters used in experiments~\cite{2013_Bloch_Harper}---$\Phi \approx 0.735 \times \pi/2$, 
and $K \approx 0.25 \times 2\pi{\rm kHz}$, 
the damping time is around $10$ms 
for an  interaction strength of $U = 0.05 \times 2\pi{\rm kHz}$~(Fig.~\ref{fig:dampingtime}).  
Thus, our predicted interaction-induced cyclotron decoherence should be observable within 
the experimental time scales for moderate values of on-site interaction strength.
In the intermediate regime $t<\tau_{\rm damp}$, we find that the time dependence of 
$P_\psi(t)$ can be empirically  described (see Fig.~\ref{fig:depletionvsdamping}) by a two-parameter fitting formula  
\be
{ P}_{\rm fit}  (t) = \frac{1}{1+ \gamma -\gamma e^{-(t/\tau)^2}}, 
\label{eq:fitfunction} 
\ee
where $\tau$ and $\gamma$ are the fitting parameters. 
This fitting formula is proposed from extending our perturbative results (to present below) to longer time. 
After the cyclotron motion relaxes, i.e., $t > \tau_{\rm damp}$ 
and $\overline{\vec{X}}(t)$ collapses to the plaquette center, 
$P_{\rm fit} (t)$ no longer captures the dynamics of $P_\psi (t)$ 
 (see Fig.~\ref{fig:damping_longtime}). 
We note that the decoherence process in Eq.~\eqref{eq:fitfunction} 
is not a simple temporal exponential relaxation phenomenon.

We have also studied the dependence of $\tau_{\rm damp}$ on the (complex) tunneling strength K and the applied magnetic flux $\Phi$. 
{We find that $\tau_{\rm damp}$ decreases with increasing $K$.} The dependence of $\tau_{\rm damp}$ on magnetic flux 
exhibits a non-monotonic {behavior},  having a minimum around $\pi/2$ (Fig.~\ref{fig:dampingtime}). 
When $\Phi$ reaches $\pi$, the spectra of 
${\cal H} ^{(0)}$ become degenerate (Fig.~\ref{fig:system}(b)) and the cyclotron dynamics changes dramatically. 
Actually even with the flux value close to $\pi$, 
 the $\psi$-mode occupation fraction $P_\psi (t)$,  as well as the oscillation amplitudes of $\overline{\vec{X}} (t)$, yield 
long-time oscillations, which we can attribute to the small energy scale in the single-particle {spectrum} near the 
degeneracy point.  
Also $P_\psi (t)$  is then no longer well-described by the empirical fit   $ P_{\rm fit}  (t)$ (Eq.~\eqref{eq:fitfunction}).

\begin{figure}[htp]
\begin{center}
\includegraphics[angle=0, width= \linewidth]{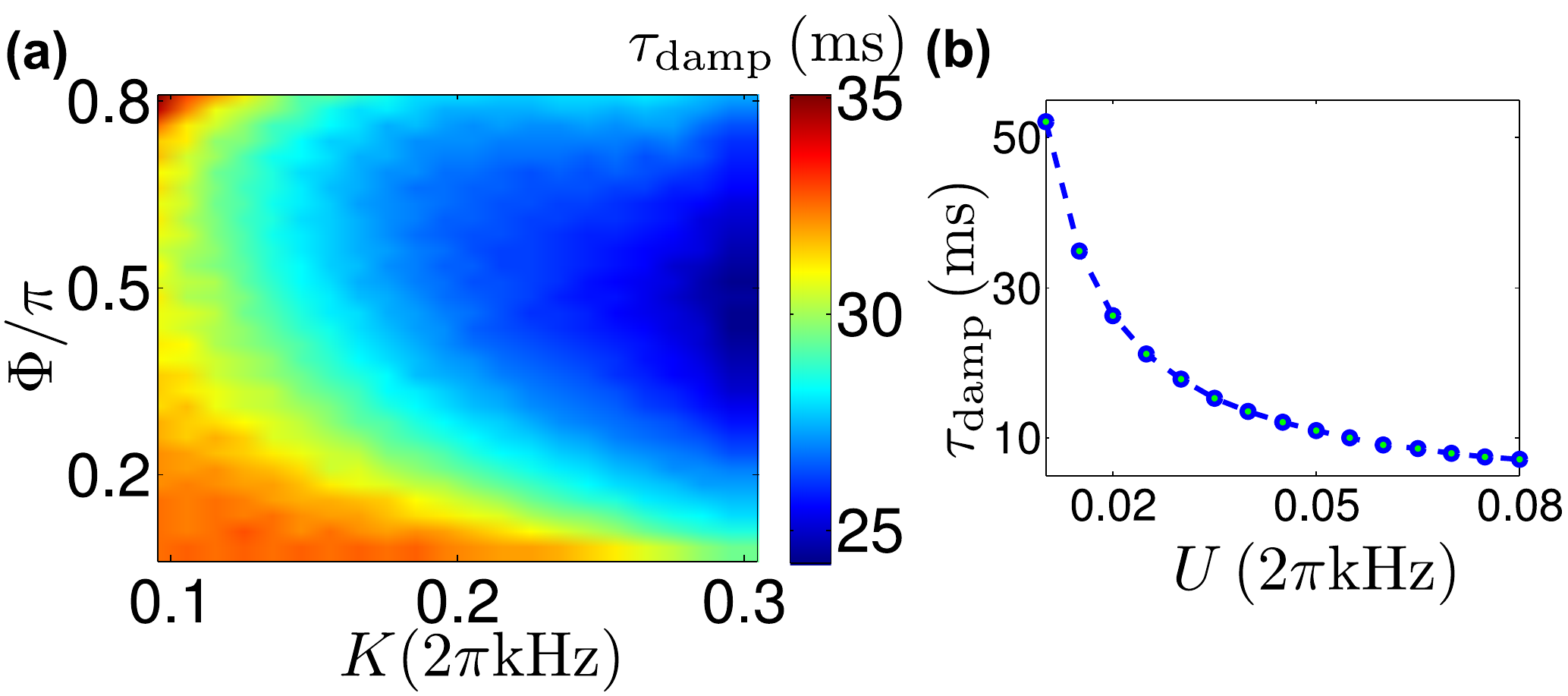}
\end{center}
\caption{Damping time with different tunnelings, interactions and magnetic 
flux. (a) The dependence of damping time $\tau_{\rm damp}$ on $K$ and $\Phi$, where $U$ is fixed 
to be $0.02 \times 2\pi$kHz. The lobe structure in (a) implies that $\tau_{\rm damp}$ decreases monotonically 
with increasing $K$ and that it has non-monotonic behavior with increasing $\Phi$. The minima of $\tau_{\rm damp}$ 
locates around $\Phi = \pi/2$.  (b) shows its dependence on $U$, where we choose $K = 0.25\times 2\pi$kHz, and 
$\Phi = 0.735 \times \pi/2$.   } 
\label{fig:dampingtime}	
\end{figure}

\begin{figure}[htp]
\begin{center}
\includegraphics[angle=0, width=.8\linewidth]{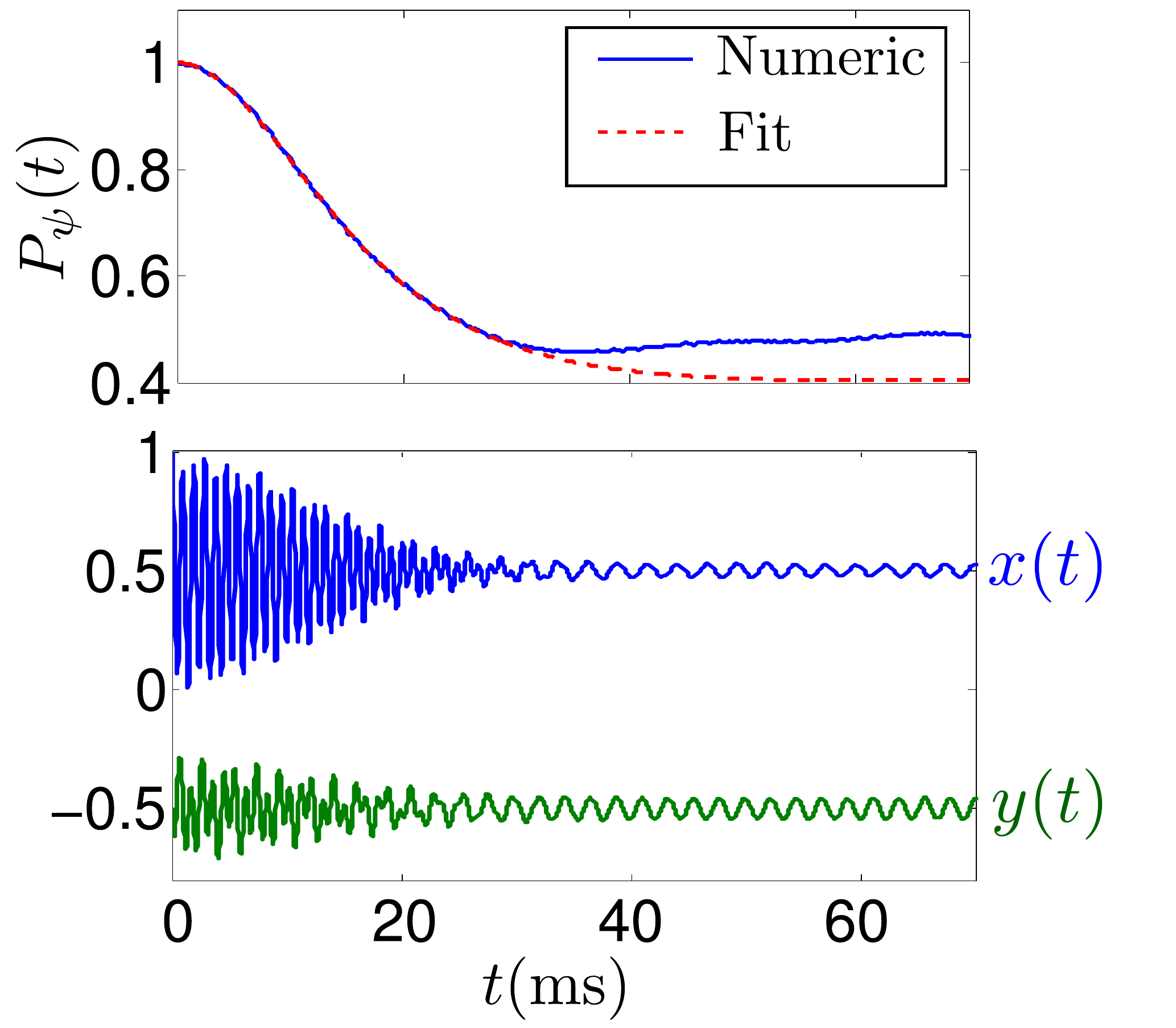}
\end{center}
\caption{Long time behavior of the cyclotron decoherence. In this plot we use $U = 0.02 \times 2\pi {\rm kHz}$, 
$K = 0.25 \times 2 \pi {\rm kHz}$ and $\Phi = 0.735 \times \pi/2$. } 
\label{fig:damping_longtime}	
\end{figure}

\subsection{Perturbative analysis}  
To better understand the cyclotron damping found in the numerics, we {carry out a} perturbative analysis with 
the standard time-dependent perturbation theory (see the Appendix).
Here it is useful to introduce single particle modes $\chi_{l = 1,2,3} ^\dag (t) |0\rangle$, which 
are orthogonal to $\psi^\dag (t) |0\rangle$. These modes $\{ \psi^\dag (t) , \chi_l^\dag (t)  \}$ form an 
instantaneous complete basis for the 
single-particle states. Similar to $\psi^\dag (t)$, we have 
$\chi_l ^\dag (t) = e^{-i H_0 t} \chi_l^\dag (0) e^{iH_0 t}  $.    
The operators $b_j^\dag $ are then expanded as 
$b_j ^\dag  = \psi_j ^* (t) \psi^\dag (t) + \chi_{lj} ^* (t)  \chi_l ^\dag  (t).  $
The occupation fraction of $\psi(t)$ is obtained as 
\bea 
&&  P_\Psi (t) =  \nn \\
 &&  1- U^2 (N-1)^2  \sum_l |I_l|^2 
- U^2 (N-1) \sum_{l_1 l_2} |I_{l_1 l_2} |^2, 
\eea 
with 
\bea 
I_l &=& \int _{t_0} ^{t} dt' \sum_{j} | \psi_{j} (t') | ^2 \psi_j (t') \chi_{lj} ^* (t'), \nn \\
I_{l_1 l_2} &=&  \int_{t_0} ^t dt' \sum_{j} \psi_{j} (t) ^2 \chi_{l_1j}^* (t') \chi_{l_2 j}^* (t').
\label{eq:integrals}  
\eea 
Expanding the single-particle wavefunctions $\psi_j (t)$ and $\chi_{lj} (t)$ in 
the eigen-basis of ${\cal H}^{(0)}$ as 
\bea 
&& \psi_j (t) = \sum_\alpha \varphi _\alpha \lambda_j ^\alpha  e^{-i \epsilon_\alpha t}, \nn \\  
&& \chi_{lj} (t) = \sum_\alpha \kappa_{l \alpha} \lambda_j ^\alpha e^{ -i \epsilon_\alpha t}, \nn 
\eea  
[$\lambda_j ^\alpha$ is the $\alpha$th eigenstate of ${\cal H} ^{(0)}$ with energy 
$\epsilon_\alpha$], 
we get 
$I_l = C^{(1)} _l  t + {\cal O} (t^0) $, 
and $I_{l_1 l_2}  = C^{(2)} _{l_1 l_2}  t + {\cal O} (t^0)$,  
with 
\bea 
C_l ^{(1)} &=&  2 \sum_{j \alpha \alpha'} |\lambda_j ^\alpha |^2 |\lambda_j ^{\alpha'} |^2 
  |\varphi_\alpha|^2 \varphi_{\alpha'} \kappa_{l \alpha'} ^* \nn \\ 
C_{l_1 l_2} ^{(2)} &=& 2 \sum_{j \alpha \alpha'} |\lambda_j ^\alpha |^2 |\lambda_j ^{\alpha'} |^2 
  \varphi_\alpha  \varphi_{\alpha'}  \kappa_{l_1 \alpha}^* \kappa_{l_2 \alpha'}^*, 
\label{eq:C12} 
\eea
provided that there is no fine-tuned degeneracy in the {spectrum} of ${\cal H} ^{(0)}$. 
Then $P_\psi(t)$ simplifies {to} 
\bea 
 && P_\psi (t) \approx 1- U^2 t^2 \nn \\
 &\times& \left[ (N-1)^2 \sum_l |C_l^{(1)}| ^2  + (N-1) \sum_{l_1 l_2} |C_{l_1 l_2} ^{(2)} |^2 \right]. 
\label{eq:occupfraction}
\eea  
This $2$nd order perturbative result is checked against exact numerics (see Fig.~\ref{fig:depletionvsdamping}). 
The fitting formula $P_{\rm fit} (t)$ (Eq.~\eqref{eq:fitfunction}) can be thought as an empirical extension of this perturbative 
result to longer time.  
{The physical picture that emerges is $|C_l ^{(1)} |^2$ describes one-particle loss rate
and $|C ^{(2)} _{l_1 l_2}|^2$ two-particle loss rate (Fig.~\ref{fig:particleproduction}). 
The damping time is estimated {from our perturbative analysis } to be 
\bea 
\tau_{\rm damp} &\propto & \frac{U^{-1} }{\sqrt{(N-1)^2 \sum_l |C_l^{(1)}| ^2  + (N-1) 
	  \sum_{l_1 l_2} |C_{l_1 l_2} ^{(2)} |^2 }}. \nn \\ 
\label{eq:taudamp}
\eea 
Carrying out the summations in Eq.~\eqref{eq:C12} numerically, we find that  two particle processes 
dominate over single particle ones, when the particle number is not {too large,} say $N <10$. 
With bosons scattered  into {the} $\chi$ modes, the depletion of $N_\psi$ causes {the} damping of 
cyclotron motion.  The dependence of $\tau_{\rm damp}$ on {tunneling, interaction,} and 
magnetic flux found in numerical simulations 
is reproduced in the perturbative analysis, and in particular, the non-monotonic dependence on 
the magnetic flux is reproduced. The long-time oscillations in $P_\psi (t)$ 
show up naturally in the integrals of  Eq.~\eqref{eq:integrals}  near 
$\pi$-flux, where the spectral degeneracy actually invalidates Eq.~\eqref{eq:C12}.

{Given the perturbative analysis, the damping phenomena in cyclotron motion are expected to be generic for interacting bosons 
in artificial magnetic fields. Despite the used specific model Hamiltonian (Eq.~\eqref{eq:Ham}) in numerical simulations, 
the described damping physics is  rather model {\it independent}.}

\begin{figure}[htp]
\begin{center}
\includegraphics[angle=0, width= \linewidth]{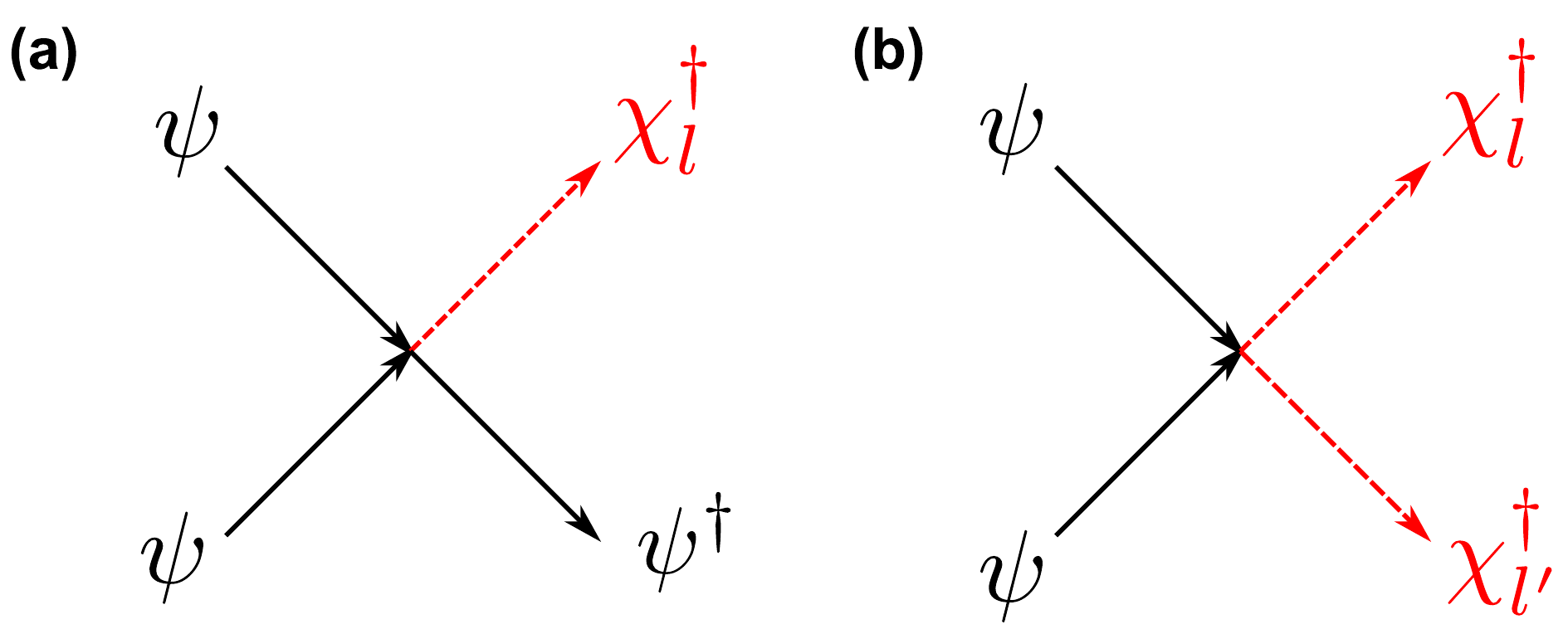}
\end{center}
\caption{Schematic diagrams of particle-loss from the initially occupied single-particle mode $\psi$. 
Bosons  are scattered into the modes $\chi_l$ in interaction processes. 
(a) and (b) illustrate the single- and two-particle loss, respectively. } 
\label{fig:particleproduction}	
\end{figure}

\section{Strong interactions and dynamical localization} 
We further look at stronger interactions, which 
are  potentially accessible in experiments, for example by implementing deep lattices. 
The perturbative analysis would no longer be reliable in the strongly interacting limit.  Our numerical simulations  show 
that bosons tend to localize for strong interaction, suppressing cyclotron motion completely. 
The quantity characterizing the localization phenomenon is the number imbalance among the four sites 
\bea 
\Delta n(t) = (n_3 (t) + n_4(t)) - (n_1(t) + n_2(t) ), 
\label{eq:nt}
\eea 
whose time average 
$$
\overline{\Delta n} = \frac{1}{T} \int_0 ^T  dt \Delta n (t) 
$$
distinguishes localized and delocalized states.  
In our simulations, we choose $T$ to be $2$ seconds and 
convergence is checked for longer time. 
As shown in Fig.~\ref{fig:localization},  
in a delocalized state with weak interactions,  the number imbalance $\Delta n$ oscillates fast (at tunneling time scale) in time and 
the time average $\overline {\Delta n}$ vanishes. In a localized state with strong interactions, 
$\Delta n$ still 
oscillates in time  but is otherwise always positive, and thus 
$\overline{\Delta n}$ 
is finite, meaning that bosons are localized on sites $3$ and $4$. The particle transfer from sites $3$ and $4$ to other two sites is  suppressed. 
An intuitive picture to understand this localization is that the tunneling probability, 
with large repulsion, is greatly suppressed because bosons have to tunnel 
all together in order to preserve energy. 
In the intermediate/crossover regime, the dynamics in $\Delta n$ yields 
fluctuations at very long time scale, which makes it challenging to determine a precise  transition point in numerics. 
Another property of the localized state is that  the $\psi$-mode occupation fraction $P_\psi (t)$ yields fast oscillations, i.e., is no longer 
quasi-static. 
This peculiar dynamical localization of strongly repulsive bosons is 
a generalization of self-trapping in double-wells to the plaquette system, 
and is an important testable prediction of our theory.

\begin{figure}[htp]
\begin{center}
\includegraphics[angle=0, width=\linewidth]{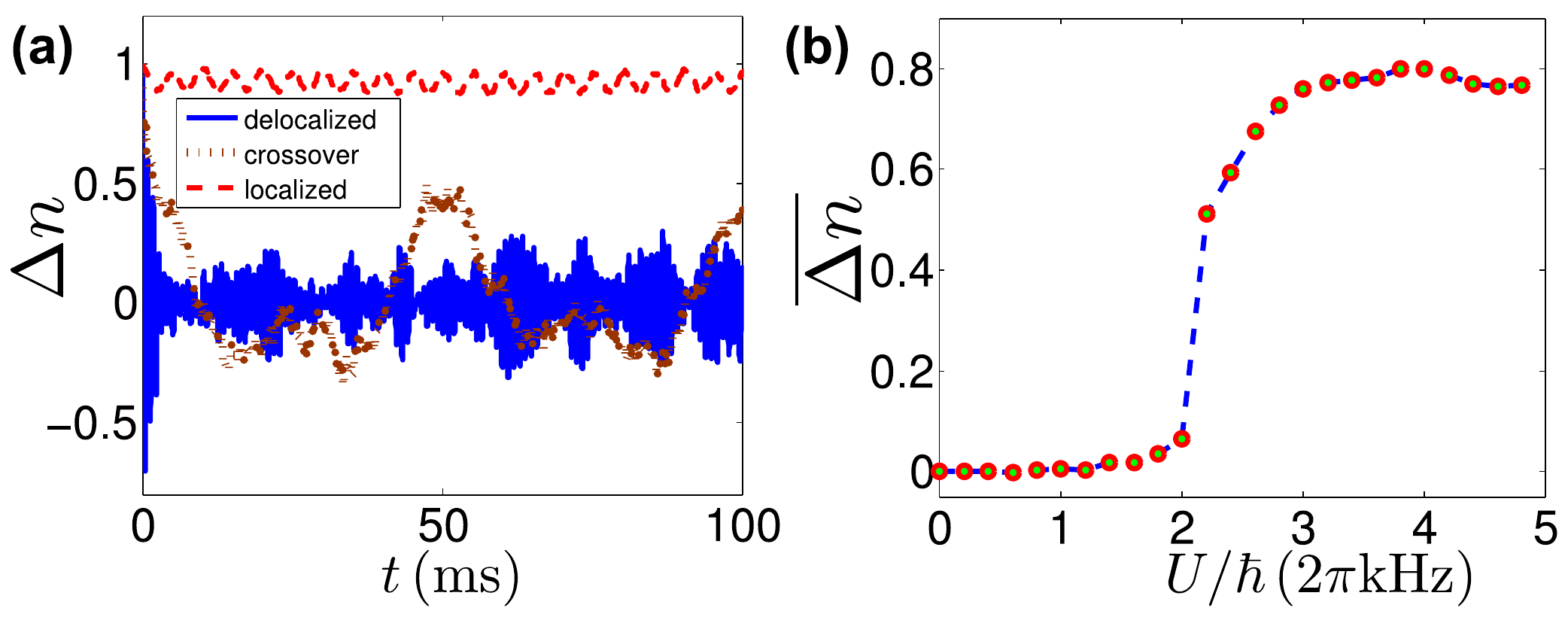}
\end{center}
\caption{Dynamical localization at strong interaction. (a), The number imbalance dynamics 
$\Delta n (t)$ (Eq.~\eqref{eq:nt}) for delocalized, localized and intermediate states, where the interaction strengths are chosen to be 
$U/\hbar = 0.2, 5, 1$ ($2\pi$ kHz), respectively. (b), Time averaged number imbalance $\overline{\Delta n}$ varying interaction strengths. 
In this plot we use $K = 0.25\times 2\pi$kHz, and 
$\Phi = 0.735 \times \pi/2$. } 
\label{fig:localization}	
\end{figure}

\section{Discussion and Conclusion} 
Although this work focused on a specific model Hamiltonian as motivated by the recent experiments~\cite{2013_Ketterle_Harper,2013_Bloch_Harper}, 
the studied interaction  induced damping in atomic cyclotron motion is expected to be a generic phenomenon. 
In particular, the damping mechanism as shown in Fig.~\ref{fig:particleproduction} and the derived damping time 
in Eq.~\eqref{eq:taudamp}  are actually model-independent and directly applicable to more generic magnetic Hamiltonians as well.   
For example, the neglected trap effects as in the experimental setup~\cite{2013_Ketterle_Harper,2013_Bloch_Harper}  could be easily included  
within our developed framework. 
The presence of the shallow trap in principle generates weak potential difference among the 
four sites (Fig.~\ref{fig:system}(a)) and further modifies the tunneling amplitudes, and  such effects 
are captured by our analytic formula (Eq.~\eqref{eq:taudamp}). With a reasonable assumption that 
the induced potential difference  and the modified tunneling amplitudes are smaller than $10\%$ of $J$, we 
find that the physics presented in this work is robust. 
One relevant question in this context is whether the damping discovered by us is 
really a `quantum collapse' phenomenon (e.g. Jaynes-Cummings model~\cite{1963_Jaynes_Cummings}) with the revival of 
the cyclotron motion at a very long time.  It is perhaps possible, in principle, for 
the system to revive at a very long time, but the fact that our analytical theory agrees 
with our direct numerical simulations and that we see no revival in the simulation 
indicates that such a revival, even if it happens, will occur at an unphysically long 
time of little interest to laboratory experiments.

Our predictions of interaction induced damping, decoherence, and dynamical localization 
(i.e. complete suppression) of the recently reported bosonic cyclotron motion~\cite{2013_Bloch_Harper} 
in optical lattices in the presence of an artificial magnetic flux should be directly experimentally 
observable since all our results presented in this work use  reasonable parameters easily achieved in 
the laboratory.  The observation of our predicted novel dynamical phenomena will be a direct 
manifestation of interaction effects on the quantum dynamics of Bose-Hubbard model in an   effective magnetic field.

\section{Acknowledgments} 
We would like to thank Jay Deep Sau, Kai Sun and Anatoli Polkovnikov for helpful  discussions. This work is supported 
by JQI-NSF-PFC,  ARO-Atomtronics-MURI, and AFOSR-JQI-MURI.

\appendix 
\section{Details of perturbative analysis for cyclotron damping} 
\label{sec:perturbdetails}
The details of perturbative analysis of the cyclotron damping dynamics are given here. 
With standard perturbation theory, the time-dependent quantum state in the interaction picture reads 
\bea 
|\Psi_I (t) \rangle = A(t) |\Psi_I ^{(0)} \rangle + | \Psi_I ^{(1)} (t) \rangle 
+  |\Psi_I^{(2)} \rangle + 
  {\cal O} (U ^3) , 
\eea 
with 
the leading part $|\Psi_I ^{(0)} \rangle  = |\Psi (t=0) \rangle$, 
the renormalization factor 
\bea 
A (t) &=&  1- i\int _{t_0} ^t d t' \langle \Psi_I ^{(0)} | V_I (t') | \Psi_I ^{(0)} \rangle \nn \\
&-&  \int_{t_0} ^t dt' \int _{t_0} ^{t'} dt{''} 
  \langle \Psi_I ^{(0)} | V_I (t') V_I (t'') | \Psi_I ^{(0)} \rangle , 
\eea 
the first order correction 
$$
 | \Psi_I ^{(1)} (t) \rangle = -i \int _{t_0} ^t dt' {\cal P} V_I(t') | \Psi_I^{(0)}\rangle, 
$$
and the second order correction 
$$
 | \Psi_I ^{(2)} (t) \rangle =- \int _{t_0} ^t dt' \int _{t_0} ^{t'} dt{''}  {\cal P} V_I(t')  V_I (t'') | \Psi_I^{(0)}\rangle, 
$$ 
where the projection operator is ${\cal P}  = 1- |\Psi_I ^{(0)} \rangle \langle \Psi_I ^{(0)} | $ and 
the interaction term $V_I (t)  = e^{iH_0 t}V e^{-iH_0 t}$. 
Then the occupation number of the $\psi(t)$ mode, $N_\psi (t)$, is given by 
\be
 N_\psi (t) = N |A(t)|^2 + \langle \Psi _I ^{(1)} (t) | \psi ^\dag \psi | \Psi_I ^{(1)} \rangle 
+ {\cal O} (U^3) . 
\ee 
{The state $|\Psi_I ^{(2) } \rangle$ does not contribute to this order because 
$\langle \Psi_I ^{ (0)}  | \psi^\dag \psi |\Psi_I ^{(2) } \rangle=0$.} 

It is useful to introduce single particle modes $\chi_{l = 1,2,3} ^\dag (t) |0\rangle$, which 
are orthogonal to $\psi^\dag (t) |0\rangle$. These modes $\{ \psi^\dag (t) , \chi_l^\dag (t)  \}$ form an 
instantaneous complete basis for the 
single-particle states. Similar to $\psi^\dag (t)$, we have 
$\chi_l ^\dag (t) = e^{-i H_0 t} \chi_l^\dag (0) e^{iH_0 t}  $,   
{and $\chi_l (0)$ will be shortened as $\chi_l $ in the following.} 
The operators $b_j^\dag $ are then expanded as 
$$b_j ^\dag  = \psi_j ^* (t) \psi^\dag (t) + \chi_{lj} ^* (t)  \chi_l ^\dag  (t).  $$ 
The renormalization factor $A(t)$ is given by 
\bea 
  1- |A(t)|^2 
&=& U^2 N(N-1)^2 \sum_l |I_l| ^2  \nn \\
&+&   \frac{1}{2} U^2 N(N-1) \sum_{l_1 l_2}  |I_{l_1 l_2} |^2, 
\eea
with $I_l$ and $I_{l_1 l_2}$ given in Eq.~\eqref{eq:integrals}. 
The perturbed state $|\Psi_I ^{(1)} (t) \rangle$ is 
\bea 
&& |\Psi _I ^{(1)} (t) \rangle 
 = -i (N-1)\sqrt{N}U  \left[ \sum_l I_l   \chi_l ^\dag \right] 
      \frac{\psi^{ \dag N-1} }{\sqrt{(N-1)!} } |0\rangle \nn \\ 
&& -\frac{i}{2} \sqrt{N(N-1)}U \left[ \sum_{l_1 l_2} I_{l_1 l_2} \chi_{l_1} ^\dag \chi_{l_2} ^\dag\right] 
	\frac{\psi^{\dag N-2} }{ \sqrt{(N-2)!} } |0\rangle . 
\eea 
The obtained occupation fraction of the $\psi(t)$ mode is given in Eq.~\eqref{eq:occupfraction}.

\bibliography{magneticboson}
\bibliographystyle{apsrev}

\end{document}